\documentclass[final,authoryear, 5p, times]{elsarticle}

\makeatletter
\def\ps@pprintTitle{%
  \let\@oddhead\@empty
  \let\@evenhead\@empty
  \def\@oddfoot{\reset@font\hfil\thepage\hfil}
  \let\@evenfoot\@oddfoot
}
\makeatother

\usepackage{graphics}
\usepackage{graphicx}

\usepackage{amsmath,amsfonts,amssymb,amsthm}
\usepackage{xcolor}
\usepackage{soul}

\usepackage{newtxtext}
\usepackage[varvw]{newtxmath}

\begin{document}

\begin{frontmatter}


\title{Interdependence of sodium and potassium gating variables in the Hodgkin-Huxley model}

\author{L\'{\i}zia Branco}
\ead{lizia.branco@kit.edu}

\author{Rui Dil\~ao}
\ead{ruidilao@tecnico.ulisboa.pt}
\address{University of Lisbon, Instituto Superior T\'ecnico, 
Av. Rovisco Pais, 1049-001 Lisbon, Portugal}

\begin{abstract}
We explore the relationship between sodium (Na$^+$) and potassium (K$^+$) gating variables in the 4-dimensional (4D) Hodgkin-Huxley (HH) electrophysiology model, and reducing its complexity by deriving new 3D and 2D models that maintain the dynamic properties of the original model. The new 3D and 2D models are grounded in the relationship $h \simeq c(I) - n$ between the gating variables $h$ and $n$ of the 4D HH model, where $c(I)$ depends of the input external stimulus, indicating an interdependence between the dynamics of Na$^+$ and K$^+$ transmembrane voltage-gated channels. The presence of Na$^+$/K$^+$-ATPase pumps along the axon may explain this interdependence. We derive the corresponding cable equations for the two new HH-type models and demonstrate that the action potential propagates along the axon at a speed given by $v(R, C_m) = \alpha / (C_m R^{\beta}):= \gamma D^{\beta}$, where $\alpha > 0$, $0 < \beta < 1$, and $\gamma$ are constants independent of the local stimulus intensity, $D$ is the diffusion coefficient of the electric signal along the axon, $C_m$ is the axon transmembrane capacitance, and $R$ is the axon conducting resistivity. 
\end{abstract}


\end{frontmatter}

\section{Introduction}\label{sec:intro}

In 1952, Hodgkin and Huxley presented a mathematical model describing the neuronal response to external electric stimuli under voltage-clamp conditions. The Hodgkin-Huxley (HH) model has significantly advanced our understanding of the mechanisms underlying the generation of neuron action potentials and their propagation along the axon. The HH original model comprises differential equations that describe the time-dependent behaviour of three ion-specific channels: voltage-gated sodium (Na$^+$) channels, voltage-gated potassium (K$^+$) channels, and leak channels. 

Despite its strong predictive power for the dynamics of ionic channels, the HH model, with its four coupled nonlinear equations and 25 parameters, hinders a clear understanding of the system dynamics and limits its efficiency in modelling neuronal networks. 
To address these challenges, we examined the specific role of each ionic channel in the HH model's dynamic behaviour and developed new reduced electrophysiological models that describe the neuronal response to external electric stimuli while preserving the model's dynamical properties.
This approach allowed us to investigate the relationship between the action potentials' spiking frequency, the electric stimulus's intensity, and various cell-specific parameters. Additionally, we explored how axon resistivity influences the speed and shape of action potential spikes.

This paper is organised as follows. In the next section, we derive and compare two reduced HH-type models based on a seemingly interdependent relationship between the sodium and potassium gating variables. Although FitzHugh first reported a relationship between these channels' gates of the form $h \simeq c - n$, where $c$ is a constant, \cite{FH}, the electrophysiological implications were seldom explored in detail, a gap we have addressed in our study.

In section~\ref{sec3}, we introduce the spatial dependence of the reduced models. We assume that the axon comprises a series of channel segments connected by junctions characterised by a resistivity parameter, \cite{End}. Subsequently, we compare the dynamic aspects of action potential propagation along the axon of the original 4D HH  model with active Na$^+$ and K$^+$ channels to those of the derived reduced 3D and 2D dynamic models. Additionally, we derive the propagation properties of action potentials as they relate to the axon parameters in response to currents originating from the soma and other external stimuli. The solitonic characteristics of certain action potential responses are discussed. Furthermore, we analyse the direction of action potential propagation along the axon when current clamp signals are introduced away from the soma. This simulates the effects of branching connections to the axon, external current stimuli, and patch clamp-type experiments. In the final section~\ref{sec4}, we summarise and discuss the main conclusions of the paper.

\section{The reduced Hodgkin-Huxley type models}\label{sec2}

Hodgkin and Huxley considered that the axon membrane of the squid {\it Loligo} has three main independent ion channels: a sodium (Na$^+$) active channel, a potassium (K$^+$) active channel, and a leak channel through which different ions could pass. By changing the composition of the extracellular space and evaluating the response of each channel's ionic current, Hodgkin and Huxley developed an empirical model where the specific values of the different parameters were chosen to fit the data best. The HH model  consists of the following four nonlinear ordinary differential equations \cite{HH1,HH2}
\begin{equation}
\begin{array}{rcl} \displaystyle
   C_m \frac{\mathrm{d} V}{\mathrm{~d} t} &=&I-\bar{g}_{\hbox{\tiny K}} n^4\left(V-V_{\hbox{\tiny K}}\right)\\ \displaystyle
	&&\displaystyle -\bar{g}_{\hbox{\tiny Na}} m^3 h\left(V-V_{\hbox{\tiny Na}}\right)-\bar{g}_{\hbox{\tiny L}}\left(V-V_{\hbox{\tiny L}}\right) \\ \displaystyle
	 \frac{\mathrm{d} n}{\mathrm{d} t} &=&\displaystyle \alpha_n(V)(1-n) - \beta_n(V)n\\ [8pt]\displaystyle
	 \frac{\mathrm{d} m}{\mathrm{d} t} &=& \displaystyle \alpha_m(V)(1-m) - \beta_m(V)m \\ [8pt]\displaystyle
        \frac{\mathrm{d} h}{\mathrm{d} t} &=&\displaystyle \alpha_h(V)(1-h) - \beta_h(V)h
\end{array}
 \label{eq-HH}
    \end{equation}
where
\begin{equation}
\begin{array}{lclrcl}\displaystyle
\alpha_n &=&\displaystyle 0.01 \phi \frac{-V+10}{e^{(-V+10)/10}-1},  &\displaystyle  \beta_n &=&\displaystyle 0.125 \phi e^{-V/80},\\ \displaystyle
\alpha_m &=&\displaystyle 0.1 \phi \frac{-V+25}{e^{(-V+25)/10}-1},  &\displaystyle  \beta_m &=&\displaystyle 4 \phi e^{-V/18},\\ \displaystyle
\alpha_h &=&\displaystyle 0.07 \phi e^{-V/20},  &\displaystyle   \beta_h &=&\displaystyle  \phi \frac{1}{e^{(-V+30)/10}+1},\\ \displaystyle
\phi &=&\displaystyle 3^{(T-6.3)/10}. &&&
\end{array}
\label{eq-HH2}
\end{equation}
In this model, $V=V_{in}-V_{out}$ is the transmembrane potential drop relative to the potential outside the cells and measured in mV, $I$ is a transmembrane current density measured in $\mu$A/cm$^2$, and time is measured in ms.  The calibrated values of the ionic Nernst potentials for the squid {\it Loligo} at a temperature $T=6.3^{\circ}$C are  $V_{\hbox{\tiny Na}} = 115$~mV, $V_{\hbox{\tiny K}} = -12$~mV and $V_{\hbox{\tiny L}} = 10.613$~mV. The transmembrane capacitance is $C_m = 1$~$\mu F/$cm$^2$. The gating variables $n$, $m$, and $h$  describe the closing and opening of the K$^+$,  Na$^+$ and leak active channels and assume values between $0$ and $1$. The parameters $\bar{g}_{\hbox{\tiny K}} = 36$~mS/cm$^2$, $\bar{g}_{\hbox{\tiny Na}}=120$~mS/cm$^2$ and $\bar{g}_{\hbox{\tiny L}} = 0.3$~mS/cm$^2$, where S=$1/\Omega$,  are the maximal potassium, sodium and leak conductance, respectively, and their values were obtained experimentally. Due to the magnitude of $\bar{g}_{\hbox{\tiny L}} $, we neglect the effect of the leak channel in the following analysis. The current term $I$ describes an external input from the soma to the axon first segment or a current source injected at any point of the axon. The HH model's current source $I$ is independent of $V$.

The HH equations \eqref{eq-HH}-\eqref{eq-HH2} have been derived assuming that the sodium and potassium voltage-gated channels are independent.

\begin{figure} 
\centering
\includegraphics[width=0.45\textwidth]{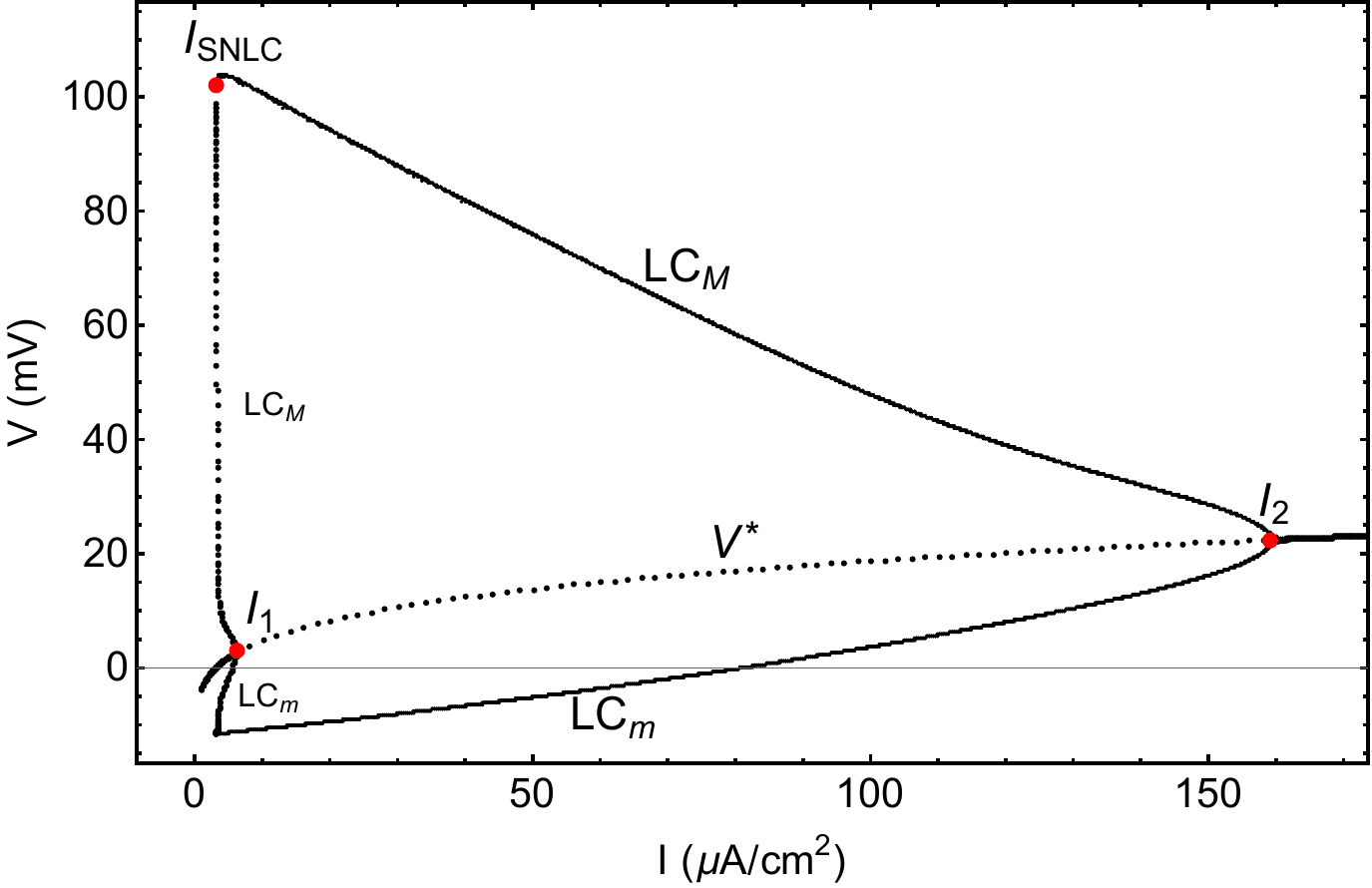}
 \caption{Bifurcation diagram of the HH equations \eqref{eq-HH}-\eqref{eq-HH2} as a function of the electric stimulus $I$ is presented. Parameter values: $C_m = 1$~$\mu F/$cm$^2$,  $V_{\hbox{\tiny Na}} = 115$~mV, $V_{\hbox{\tiny K}} = -12$~mV, $\bar{g}_{\hbox{\tiny K}} = 36$~mS/cm$^2$, $\bar{g}_{\hbox{\tiny Na}}=120$~mS/cm$^2$, $\bar{g}_{\hbox{\tiny L}} = 0.0$~mS/cm$^2$, $I_1 \approx 6.18$~$\mu$A/$\text{cm}^2$, $I_2 \approx 159.20$~$\mu$A/$\text{cm}^2$ and $I_{\hbox{\tiny SNLC}} \approx 3.15$~$\mu$A/$\text{cm} ^2$. Dotted lines correspond to unstable states of the HH model equations, while continuous lines represent stable states. Action potential solutions of the HH equations are obtained for $I_{\hbox{\tiny th}}\le I\le I_1$, where $I_{\hbox{\tiny th}}\simeq 2.25$~$\mu$A/$\text{cm}^2$. For $I_1< I<  I_2$, we observe spiking solutions.}
 \label{fig:fig1}
\end{figure}

The system of equations \eqref{eq-HH}-\eqref{eq-HH2} has one fixed point with coordinates $p=(V^{*}, n^{*}, m^{*},h^{*})$ that can be stable or unstable. Using the bifurcation analysis software XPPAUT (\cite{Erm}), in Figure~\ref{fig:fig1}, we show the steady state $V^{*}$ as a function of the constant electric stimulus $I$. The two points where the fixed point changes stability correspond to Hopf bifurcations: the first occurring subcritically at $I_1 \approx 6.18$~$\mu$A/$\text{cm}^2$, while the second occurs supercritically at $I_2 \approx 159.20$~$\mu$A/$\text{cm}^2$. A limit cycle (LC) is formed at these bifurcation points. In Figure~\ref{fig:fig1}, the dotted LC line corresponds to an unstable limit cycle, while the solid LC line corresponds to a stable limit cycle. The point at which these two types of limit cycles coalesce is known as a saddle-node bifurcation of limit cycles (SNLC), occurring at $I_{\hbox{\tiny SNLC}} \approx 3.15$~$\mu$A/$\text{cm}^2$. The maximum and minimum values of the membrane potential along the limit cycles are labelled as $LC_M$ and $LC_m$, respectively.

Along the first region where $p$ is stable, $I<I_{1}$, for $0\le I<I_{\hbox{\tiny th}}\simeq 2.25$~$\mu$A/$\text{cm}^2$, where $I_{\hbox{\tiny th}}$ is a threshold parameter, all the solutions of the HH model converge for the stable steady state $p$. For $I_{\hbox{\tiny th}}<I<I_1$, the solution of equations \eqref{eq-HH}-\eqref{eq-HH2} may show the intermittent firing of a finite number of action potential spikes. This behaviour is due to a type I intermittency phenomenon occurring near the SNLC bifurcation, and the exact number $M$ of action potential spikes is given by $\ln M=C-2\ln(I_{\hbox{\tiny SNLC}} - I)$, where $C$ is a constant. For $I_1 < I < I_2$, the solution of the HH equation shows a persistent spiking phenomenon in response to the stimulus intensities $I$. For $I>I_{2}$, $p$ is a stable fixed point (\cite{CD1}).
	
As noted by different authors, the profile of the solution $h(t)$ of equations \eqref{eq-HH}-\eqref{eq-HH2} approximately mirrors $n(t)$, \cite[p. 456]{FH}, \cite{Rin2}, \cite[p. 213]{KS} and \cite{Wan}. 
This property of the solutions suggests the existence of a correlation between the conductivities of the Na$^+$ and the K$^+$ channels that Hodgkin and Huxley did not originally consider. Replacing $h$ by $c-n$ in equations \eqref{eq-HH}-\eqref{eq-HH2}, where $c$ is some constant, we obtain the reduced 3D HH model 
\begin{equation}
\begin{array}{rcl}
C_m\frac{dV}{dt} &=& I - g_{\hbox{\tiny K}} n^4(V-V_{\hbox{\tiny K}}) - g_{\hbox{\tiny Na}}m^3(c-n)(V-V_{\hbox{\tiny Na}})\\[8pt] \displaystyle
\frac{dn}{dt}  &=& \displaystyle  \alpha_n(V)(1-n)-\beta_n(V)n\\[8pt]  \displaystyle
\frac{dm}{dt}  &=& \displaystyle  \alpha_m(V)(1-m)-\beta_m(V)m.
\end{array}\label{eq:3dmodel}
\end{equation}

\begin{figure}[!htb]
\centering
\includegraphics[width=0.45\textwidth]{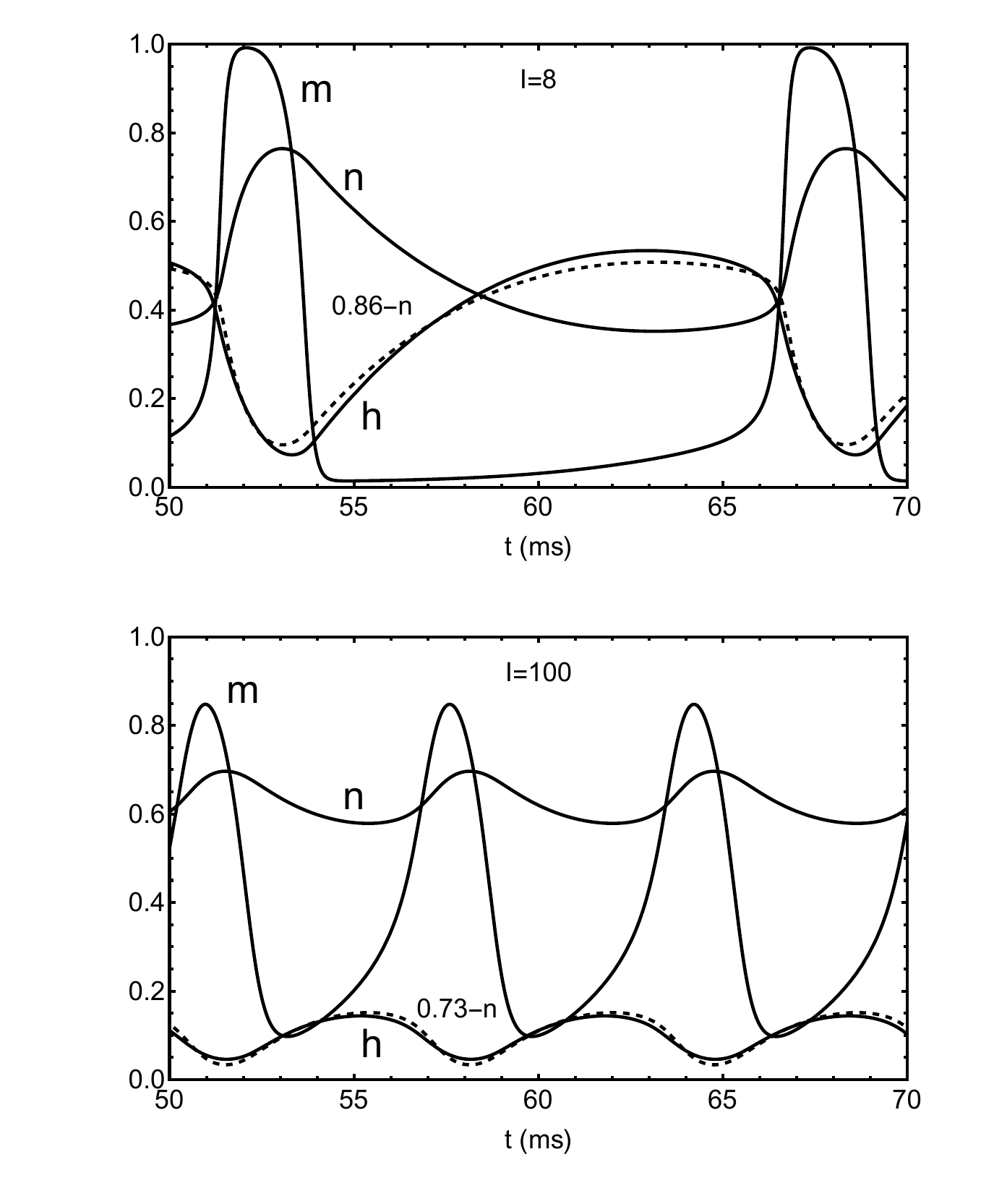}
  \caption{Relationship between $n$ and $h$ for the spiking regimes $I=8$ and $I=100$, obtained with HH equations \eqref{eq-HH}-\eqref{eq-HH2}. For $I=8$, the best fit of $h$ to $h=c_{\hbox{\tiny 3D}}(I)-n$ is for the choice $c=0.86$. For $I=100$ we obtained   $c=0.73$. The other parameters of the simulations are the same as in Figure \ref{fig:fig1}. For other choices of $I$, $c_{\hbox{\tiny 3D}}(I)$ is well described by \eqref{parc}. The gating variables $n$, $m$, and $h$ are adimensional and take values in the interval $[0,1]$.}
  \label{fig:fig2}
\end{figure}

To analyse the similarities between the solutions of equations \eqref{eq-HH}-\eqref{eq-HH2} and \eqref{eq:3dmodel}, we calculated the values of $c$ that best approximate the function $h=c-n$, obtained with the HH model for several values of $I$ (Figure \ref{fig:fig2}). A simple estimate with least squares fitting shows that the 3D reduced HH model \eqref{eq:3dmodel} well approximates the solutions of equations \eqref{eq-HH}-\eqref{eq-HH2} for
\begin{equation}
c=c_{\hbox{\tiny 3D}}(I)= \left\{
\begin{array}{ll}
1.0\ \ &\hbox{for}\ \ I\le 1\\
1.0\, I^{-0.0674}\ \ &\hbox{for}\ \ I>1.
\end{array}\right.
\label{parc}
\end{equation}
showing the dependendence of $c$ on the input current $I$.

\begin{figure}[!htb]
\centering
\includegraphics[width=0.45\textwidth]{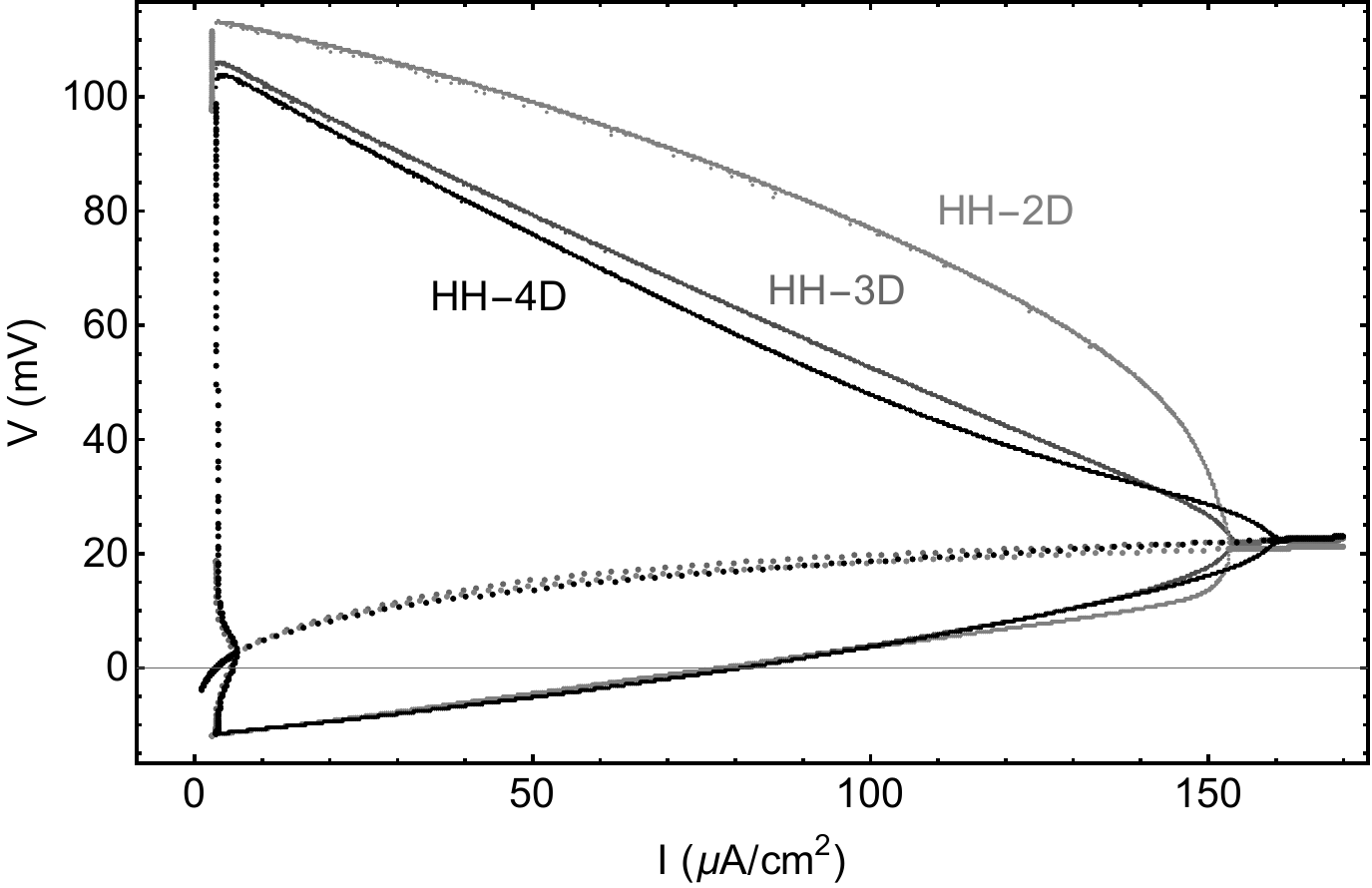}
  \caption{Bifurcation diagrams of the 2D (light grey), 3D (dark grey) and 4D (black) HH type models. In the 3D and 2D models \eqref{eq:3dmodel} and \eqref{eq:2dmodel}, respectively,  the parameter $c=c(I)$  is given by \eqref{parc}  and \eqref{parc2}. The overall behaviour of the bifurcation diagrams has the structure of a codimension 2 Bautin bifurcation scenario.}
  \label{fig:fig3}
\end{figure}

Furthermore, we investigated the impact of approximating the gating variables $n$ and $m$ by their steady-state values on the system's behaviour. While approximating $n$ by $n_{\infty}$ --- the solution of the equation $\frac{dn}{dt}=0$ --- led to models without Hopf bifurcations and periodic cycles, and all the fixed points are stable (\cite{Bra}), more realistic results are obtained by approximating $m$ by $m_{\infty}$ --- the solution of the equation $\frac{dm}{dt}=0$. 
The steady-state solution $m(t)$ of the HH equation is closely approximated in both amplitude and period by $m_{\infty}(V(t))$.
Therefore, we further simplify the reduced HH 3D model \eqref{eq:3dmodel} to the reduced 2D model
\begin{equation}
\begin{array}{rcl}
C_m\frac{dV}{dt} &=& I - g_{\hbox{\tiny K}}n^4(V-V_{\hbox{\tiny K}}) \\[8pt] \displaystyle
&&- g_{\hbox{\tiny Na}}m_{\infty}^3(c_{\hbox{\tiny 2D}}(I)-n)(V-V_{\hbox{\tiny Na}})\\[8pt] \displaystyle
\frac{dn}{dt}  &=& \displaystyle  \alpha_n(V)(1-n)-\beta_n(V)n\\[8pt]  \displaystyle
m_{\infty}(V) &=&\displaystyle  \frac{\alpha_m(V)}{\alpha_m(V)+\beta_m(V)},
\end{array}\label{eq:2dmodel}
\end{equation}
where
\begin{equation}
c_{\hbox{\tiny 2D}}(I)= \left\{
\begin{array}{ll}
1.0\ \ &\hbox{for}\ \ I\le 1\\
1.0\, I^{-0.078}\ \ &\hbox{for}\ \ I>1,
\end{array}\right.
\label{parc2}
\end{equation}
and these values for $c_{\hbox{\tiny 2D}}(I)$ were fitted in such a way that the bifurcation diagrams of the 4D and 2D models were as similar as possible with close bifurcation values (Figure \ref{fig:fig3}). 

Both model \eqref{eq:3dmodel} and \eqref{eq:2dmodel} are derived from the original HH model \eqref{eq-HH}-\eqref{eq-HH2}, and as we show below, they share the same dynamical properties.

Figure \ref{fig:fig3} depicts the bifurcation diagrams of the three model equations \eqref{eq-HH}-\eqref{eq-HH2}, \eqref{eq:3dmodel} and \eqref{eq:2dmodel}, with $c\equiv c(I)$ given by \eqref{parc} and \eqref{parc2}, respectively. 
The similarities between the bifurcation diagrams of the HH model and the reduced models where the gating variable $h$ was replaced by $c(I)-n$ suggest that the behaviour of the Na$^+$ channels can be described by a combination of the gating variables $m$ and $n$. FitzHugh and Rinzel recognised this feature. While FitzHugh considered $c=0.85$ and Rinzel $c=1$, independently of the current $I$, our analysis shows that $c(I)$ is current dependent and is better approximated by \eqref{parc} or \eqref{parc2}. Both the 2D and 3D reduced models (equations \eqref{eq:3dmodel} and \eqref{eq:2dmodel}) preserve the global features of the HH 4D (local) model while reducing its complexity. 
The bifurcation parameters for the 2D and the 3D HH reduced models are: (2D Model) $I_{\hbox{\tiny SNLC}}=2.61$, $I_1=5.31$ and $I_2=153.13$;  (3D Model) $I_{\hbox{\tiny SNLC}}=3.16$, $I_1=5.60$ and $I_2=153.32$. 

We have compared the periods of the action potential spikes in the oscillatory regions $[I_1,I_2-\varepsilon ]$ of the 4D, 3D and 2D models, where $\varepsilon=0.02$. For these cases, the periods are well-fitted by the functions
$$\begin{array}{lcl}
 per_{\hbox{\tiny 4D}}(I)&=&32.96\times I^{-0.35}\ \hbox{ms}\\
 per_{\hbox{\tiny 3D}}(I)&=&34.23\times I^{-0.37}\ \hbox{ms} \\
 per_{\hbox{\tiny 2D}}(I)&=&37.33\times I^{-0.49}\ \hbox{ms}.
 \end{array}
$$

In the regions $[I_2-\varepsilon_1 , I_2]$, with $\varepsilon_1\simeq 10^{-4}$, the periods converge to zero as $I\to I_2$ due to the existence of canard-type solutions of the three models, \cite{Bra}.
 
The HH model and the 3D and 2D reduced models have a bifurcation structure described by a Bautin bifurcation scenario, \cite{CD1} and \cite{Izh}.

This analysis shows that the transmembrane active transport of sodium and potassium is interdependent, justified by the relationship between $h$ and $c(I)-n$. This interdependence may be partially explained by the ubiquitous Na$^+$/K$^+$-ATPase, an enzyme that maintains the balance of sodium and potassium ions in the living cell. This active channel was not originally considered in the HH model, being discovered only a few years later \cite{Sko}. However, to validate this hypothesis, a comprehensive analysis integrating focused biological experiments and mathematical modelling is necessary.

\section{Results}\label{sec3}

\begin{figure}[!ht]
  \centering
  \includegraphics[width=0.5\textwidth]{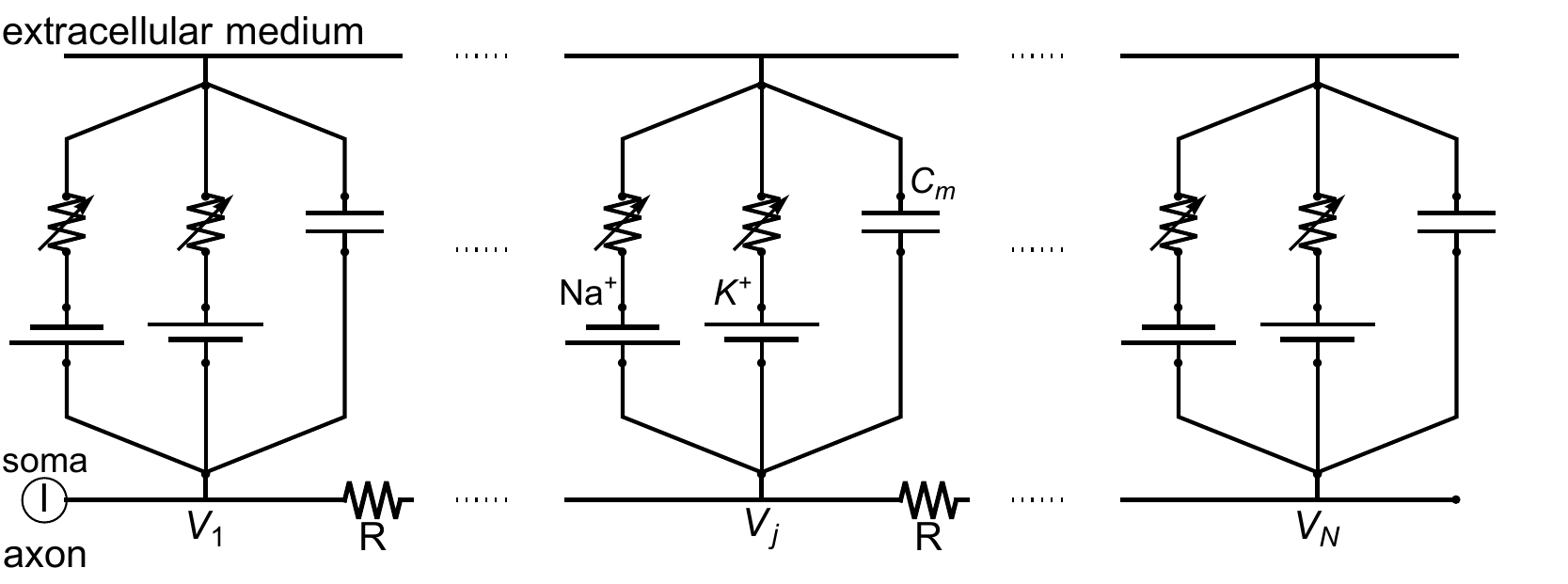}
  \caption{Electric analogue of an axon according to the Hodgkin-Huxley view, excluding leak channels (\cite{HH2}). Each group of active channels communicates with the extracellular space through Na$^+$ and K$^+$ active channels. The signal from the soma is simulated as a current source $I$, and communication between active channels is characterised by the resistance $R$. The membrane potential at each pair of active channels is $V_j$, and the extracellular space potential $V_{\hbox{\tiny out}}$ is assumed constant. The variable resistances of the channels are the inverse of the variable conductivities ${\bar g}_{\hbox{\tiny Na}}$ and ${\bar g}_{\hbox{\tiny K}}$.}
  \label{fig:fig4}
\end{figure}

The axon is a sequence of segments containing voltage-gated sodium and potassium channels. Each segment is connected to the neighbouring segments through junctions characterised by a resistivity parameter $R$. The current stimulus $I$ from the soma of a neuron is transmitted to the first node of its axon, propagating along the sequence of contiguous segments. In Figure \ref{fig:fig4}, we schematically illustrate the electric analogue of an axon for the case where it receives the stimulus from the soma.

Each pair of active channels of the axon is characterised by the potential $V_j$, and the current that flows to adjacent channels obeys Ohm's law. Under these conditions, and taking as local dynamics equations of the reduced 2D model \eqref{eq:2dmodel}, the equations describing signal propagation along the axon are
\begin{equation}
	\begin{array}{rcl}\displaystyle
	C_m \frac{dV_1}{dt} &=&\displaystyle I + \bigg(\frac{V_2-V_1}{R}\bigg) - I_{ion}(V_1, n_1,m_{\infty}(V_1),c_{\hbox{\tiny 2D}}(I))\\ [8pt]\displaystyle
	C_m \frac{dV_j}{dt} &=&\displaystyle \bigg(\frac{V_{j-1}-V_{j}}{R}\bigg)+ \bigg(\frac{V_{j+1}-V_{j}}{R}\bigg)\\ [8pt]\displaystyle
	&&  -I_{ions}(V_j, n_j,m_{\infty}(V_j),c_{\hbox{\tiny 2D}}(0)),\ \ 1<j<N \\ [8pt]\displaystyle
	C_m \frac{dV_N}{dt} &=&\displaystyle \bigg(\frac{V_{N-1}-V_{N}}{R}\bigg) - I_{ions}(V_N, n_N,m_{\infty}(V_N),c_{\hbox{\tiny 2D}}(0))\\ [8pt]\displaystyle
	\frac{dn_j}{dt} &=&\displaystyle \alpha_n(V_j)(1-n_j)-\beta_n(V_j)n_j ,\ \ 1\le j\le N\\ [8pt]\displaystyle
	m_{\infty}(V) &=&\displaystyle  \frac{\alpha_m(V)}{\alpha_m(V)+\beta_m(V)} ,\ \ 1\le j\le N\\ [8pt]\displaystyle
	I_{ions} &=&\displaystyle g_{\hbox{\tiny K}}n_j^4(V_j-V_{\hbox{\tiny K}}) \\  [8pt]\displaystyle
	&&+ g_{\hbox{\tiny Na}}m_{\infty}(V_j) ^3(c_{\hbox{\tiny 2D}}(I)-n_j)(V_j-V_{\hbox{\tiny Na}}) ,
\end{array}
\label{propagation}
\end{equation}
where $c_{\hbox{\tiny 2D}}(I)$ is defined in \eqref{parc2}, and the $\alpha$ and $\beta$ functions are the usual HH functions defined in \eqref{eq-HH2}. 
The axon has length $L=N\ell$, where $\ell$ is the length of each voltage-gated element of the axon. The diffusion coefficient is defined as $D=\ell^2/R$ with physical dimensions $S=1/\Omega$. The soma is at  $x=0$, and the presynaptic terminal is at $x=L$.


Equation \eqref{propagation} has been derived from the 4D HH model with the explicit introduction of the empirical fact that sodium and potassium voltage-gated channels are not dynamically independent, as shown by the relationship between $h$ and $(c(I)-n)$. 
In the following subsections, we compare the solutions of three models (4D, 3D and 2D) and derive their propagation properties as a function of the axon resistivity $R$ and transmembrane conductivity $C_m$.

\subsection{Signal transmission along the axon -- 4D model}

\begin{figure}
\centering
\includegraphics[width=0.35\textwidth]{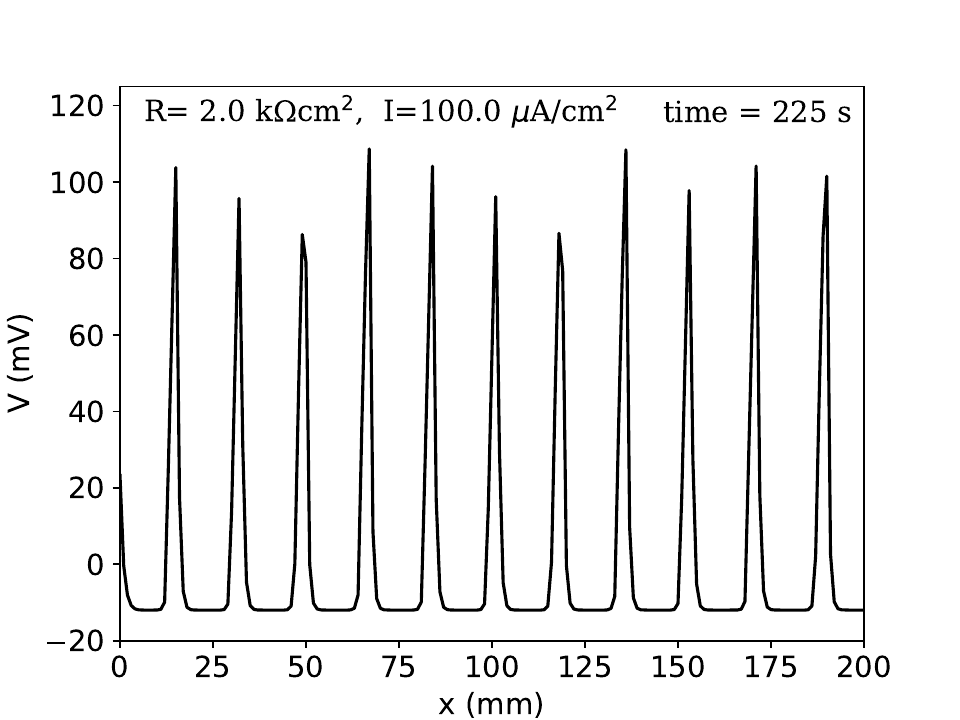}
\includegraphics[width=0.35\textwidth]{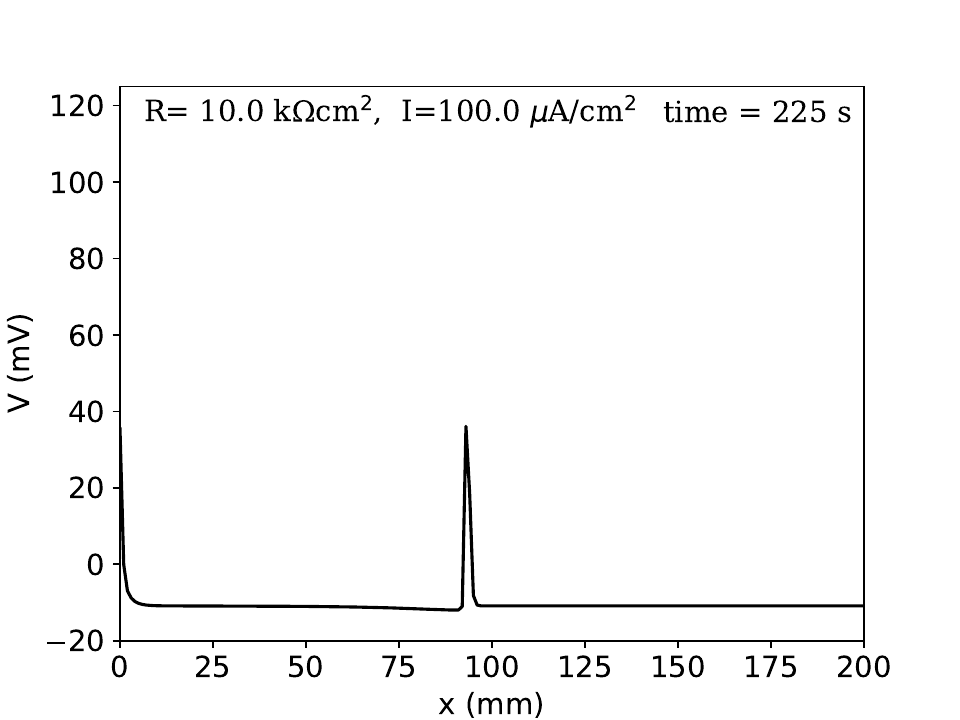}
 \caption{Transmembrane potential along an axon at time $t=225$~ms, for a soma  constant  stimulus $I=100$~$\mu$A/cm$^2$, described by the 4D HH model \eqref{eq-HH}-\eqref{eq-HH2}-\eqref{propagation}, with $\bar{g}_L =0$. In the $R=2.0$~k$\Omega$cm$^2$ case, the action potential propagates along the axon as a periodic or quasi-periodic time function.
 For  $R=10.0$~k$\Omega$cm$^2$, a solitary spiky signal forms and propagates along the axon, and after colliding with the presynaptic region of the axon,  a constant steady state is established, and no further axonal activity is observed.}
 \label{fig:fig5}
\end{figure}

The signals generated by the current $I$ that arrives from the soma travel unidirectionally along the axon. For later comparison and reference, we use the 4D HH  model \eqref{eq-HH}-\eqref{eq-HH2}, with $\bar{g}_L =0$, with diffusion constant ($D=\ell^2/R$) and boundary conditions calculated as in model equations \eqref{propagation}.  Figure~\ref{fig:fig5} depicts two action potential signals solutions along the axon at time $t=225$~ms, for a soma  ($x=0$) constant stimulus $I=100$$\mu$A/cm$^2$, and several values of the intracellular resistance $R$, measured in units of k$\Omega$cm$^2$.  We have considered an axon with $N=200$ axon segments, each with length $\ell =1$~mm. The equations \eqref{propagation} were integrated with the Euler method with $dt=0.001$~ms. The initial conditions at the steady state  $I=0$ are $$(V^*,n^*,m^*,h^*)=(-10.8781, 0.1710, 0.0138, 0.8796).$$

For $R\in [0.01,20]$, the action potential response has different dynamical properties. For $R\lnapprox 0.01$~k$\Omega$cm$^2$ or $R\gnapprox 18$~k$\Omega$cm$^2$, the axon remains at a steady state, and there is no propagation of the action potential. For $0.01\lnapprox R\lnapprox0.02$ or $4.0\lnapprox R \lnapprox 17.5$, a solitary action potential spike is produced,  propagating along the axon before the axons settle into a steady state. For $0.02\lnapprox R\lnapprox 3.9$, periodic trains of action potential are established, propagating along the axon.

\begin{figure}
\centering
\includegraphics[width=0.38\textwidth]{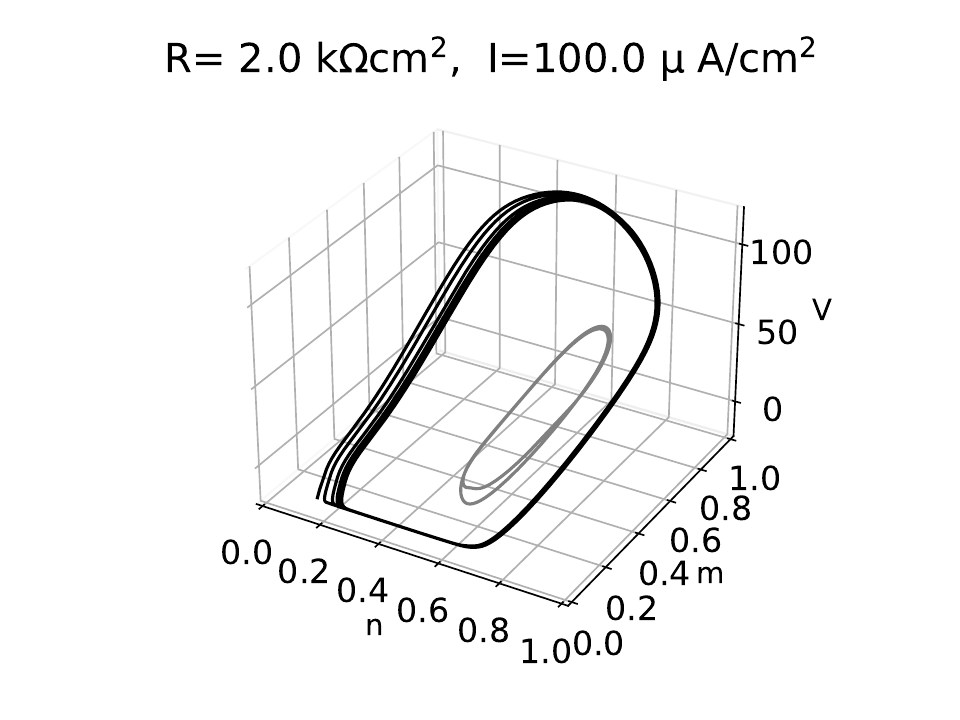}
\caption{Projection of the phase space orbits of the 4D HH model \eqref{eq-HH}-\eqref{eq-HH2}-\eqref{propagation}, with $\bar{g}_L =0$, at the soma  ($N=0$, in grey) and the middle of the axon ($N=100$, in black), during the time interval $[500,2000]$~ms.}
 \label{fig:fig6}
\end{figure}

In Figure~\ref{fig:fig6}, we show two three-dimensional phase space projections of the asymptotic orbit of the solutions of the 4D HH model, with $R=2.0$~k$\Omega$cm$^2$ and $I=100$$\mu$A/cm$^2$,  at the soma and the middle of the axon.

\begin{figure}
\centering
\includegraphics[width=0.3\textwidth]{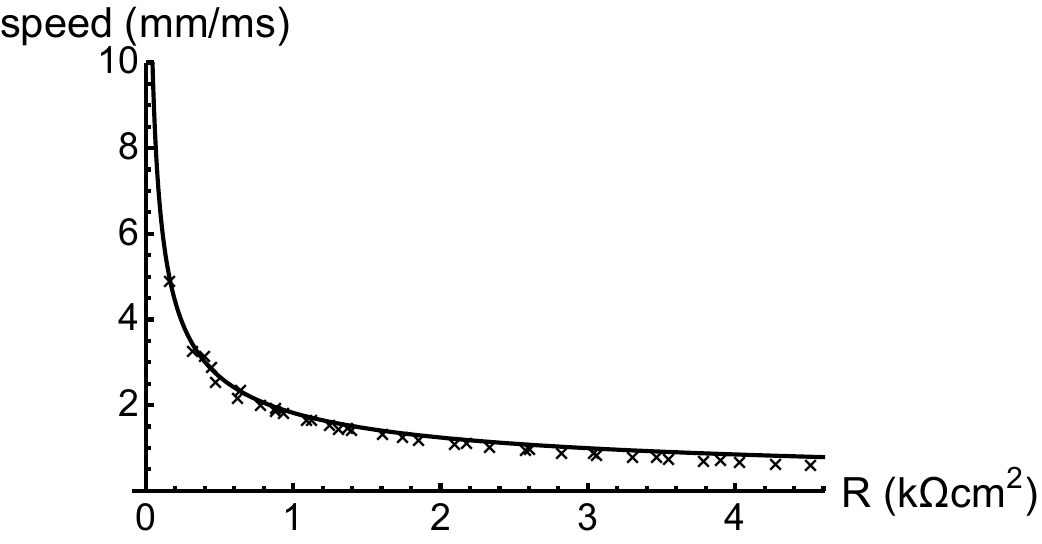}
\caption{Propagation speed of the first action potential spike as a function of the axon resistivity $R$, for $I=20,100, 140$~$\mu$A/cm$^2$ and $C_m = 1$~$\mu F/$cm$^2$, calculated from the 4D HH model (1)-(2). The action potential speed is well approximated by equation \eqref{speed2-4D}. The fitted functions \eqref{speed-4D} are practically indistinguishable from \eqref{speed2-4D}.}
 \label{fig:fig7}
\end{figure}

If an action potential spiky signal is produced, the velocity of the first spike propagates along the axon with a constant speed, \cite{CD2}. The propagation speed of the first action potential spike has been calculated numerically for several values of the soma stimulus $I$ as a function of the axon resistivity $R$. The propagation velocity depends on the resistance along the axon (Figure~\ref{fig:fig7}) but weakly depends on the stimulus $I$ at the soma. The speed  of the first action potential spike is well described by the functions
\begin{equation}
\begin{array}{lcl}
v_{\hbox{\tiny 4D}}(R,I=20)=1.83\ R^{-0.63}, \ R\in [0.4,4.5]\\
v_{\hbox{\tiny 4D}}(R,I=100)=1.81\ R^{-0.54}, \ R\in [0.02,3.9]\\
v_{\hbox{\tiny 4D}}(R,I=140)=1.82\ R^{-0.54}, \ R\in [0.01,1.4],
\end{array}
\label{speed-4D}
\end{equation}
with units of mm/ms.

This suggests that the propagation speed of the action potential along the axon is approximately described by the function
\begin{equation}
v_{\hbox{{\tiny 4D}}}(R)=\alpha \frac{1}{R^{\beta}},
\label{speed2-4D}
\end{equation}
where $\alpha=1.82$ and $\beta=0.55$. As $1/R\sim D$, we have  $v\sim D^{0.55}$, where $D$ is the diffusion coefficient.

For $I=20$~$\mu$A/cm$^2$, oscillatory responses along the axon are in the resistance interval $0.4\lnapprox R\lnapprox 10.0$. For $I=140$~$\mu$A/cm$^2$, we obtained oscillatory responses for   $R\in [0.01, 1.4]$.

\begin{figure}
\centering
\includegraphics[width=0.30\textwidth]{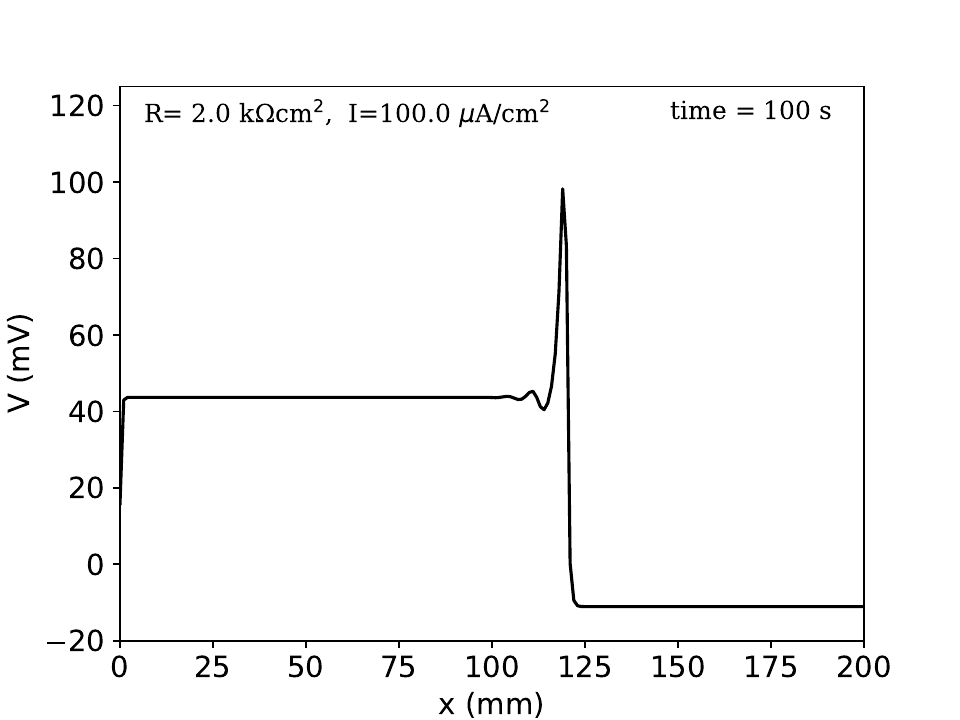}
\caption{Action potential soliton transient solution for the 3D HH modified model.}
 \label{fig:fig8}
\end{figure}

\begin{figure*}
\centering
\includegraphics[width=0.8\textwidth]{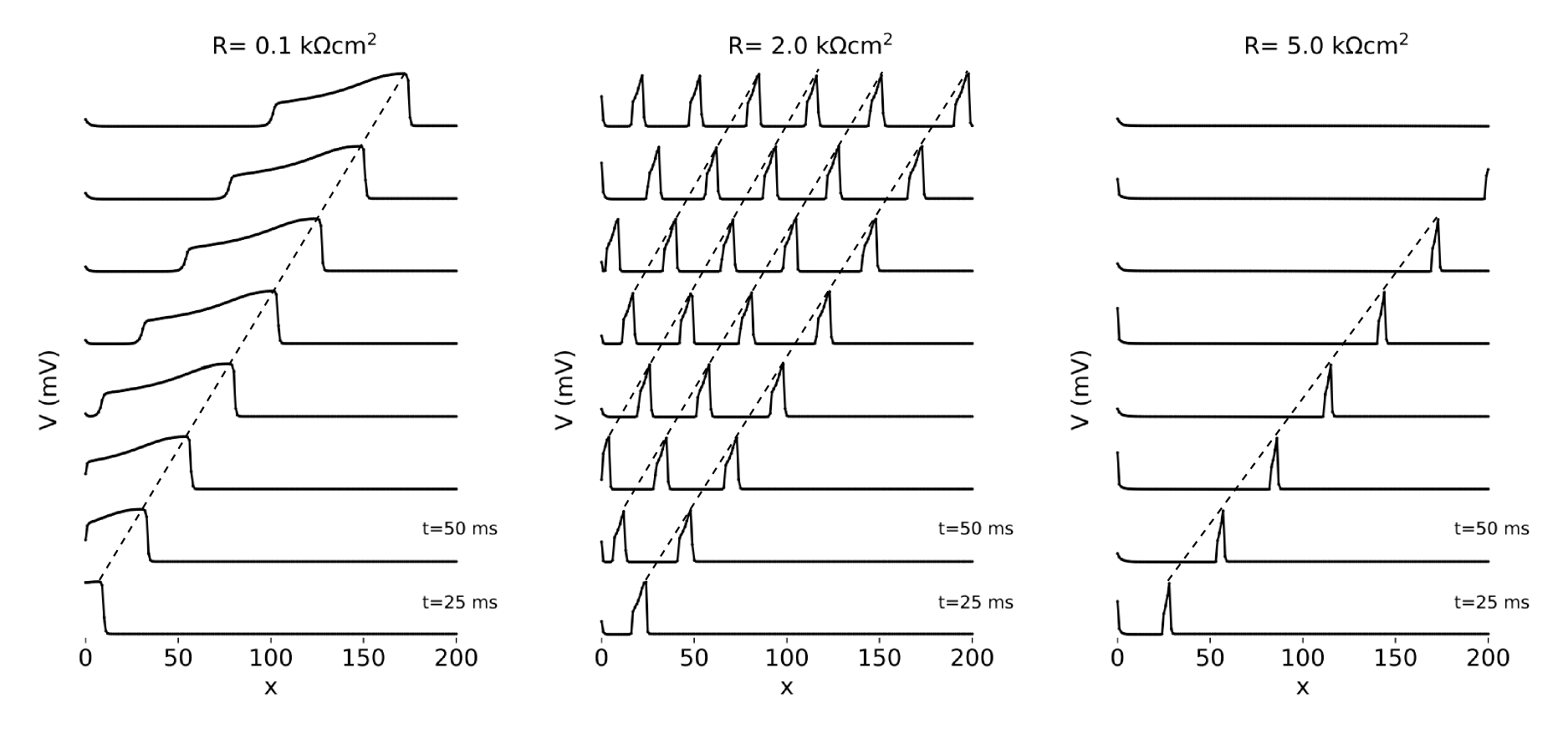}
 \caption{Propagation of the transmembrane potential along the axon with a constant stimulus at the soma with $I=100$~$\mu$A/cm$^2$ and $C_m = 1$~$\mu F/$cm$^2$, for several values of the resistance $R$, calculated with model equations \eqref{propagation} relative to the reduced 2D HH model. The vertical axis range is $[-20,125]$~mV. The axon has $N=200$ segments with lengths $\ell=1$~mm. The dashed lines are the characteristic curves of the action potential spikes, showing that the propagation speed of each spike is constant. For $R=5.0$~k$\Omega$cm$^2$, we observe a soliton-type solution that annihilates at the synaptic region of the axon and for $t>200$~ms, the axon stays in a steady state.}
 \label{fig:fig9}
\end{figure*}

\subsection{Signal transmission along the axon -- 3D model}

The same approach followed for the 4D HH model with $\bar{g}_L =0$ can be tested for the 3D model with $h=c_{\hbox{\tiny 3D}}(I)-n$, and $c_{\hbox{\tiny 3D}}(I)$ as defined in \eqref{parc}. The bifurcation diagrams for the local 4D and 3D systems are very similar. However, numerical results with the discrete setting \eqref{propagation} give soliton-type solutions with one spiky signal and the establishment of a constant steady state along the axon for $0.1\lnapprox R\lnapprox 7.0$ (Figure~\ref{fig:fig8}). In this interval, the speed of the action potential soliton is well described by the function $v_{\hbox{\tiny 3D}}=1.81/R^{0.52}$~mm/ms, which is close to the estimate for the 4D model \eqref{speed2-4D}.

\subsection{Signal transmission along the axon -- 2D model}

We now consider the reduced 2D HH model  \eqref{propagation}, where $c_{\hbox{\tiny 2D}}(I)$ is defined in \eqref{parc2}, and the $\alpha$ and $\beta$ are the usual HH functions as defined in \eqref{eq-HH2}.  

We consider  the same parameter values of the 4D model but with initial conditions $(V^*,n^*)=(-10.9506, 0.1702)$. In the 2D HH reduced model, for $0.01\lnapprox R\lnapprox 4.6$, the system propagates trains of action potentials and is periodic in time for any longitudinal position along the axon. For $R\lnapprox 0.05$ and $R\gnapprox 4.6$, one spike is produced, and, as time passes, the spatial solution along the axon converges to a steady state (Figure~\ref{fig:fig9}). 

\begin{figure}
\centering
\includegraphics[width=0.23\textwidth]{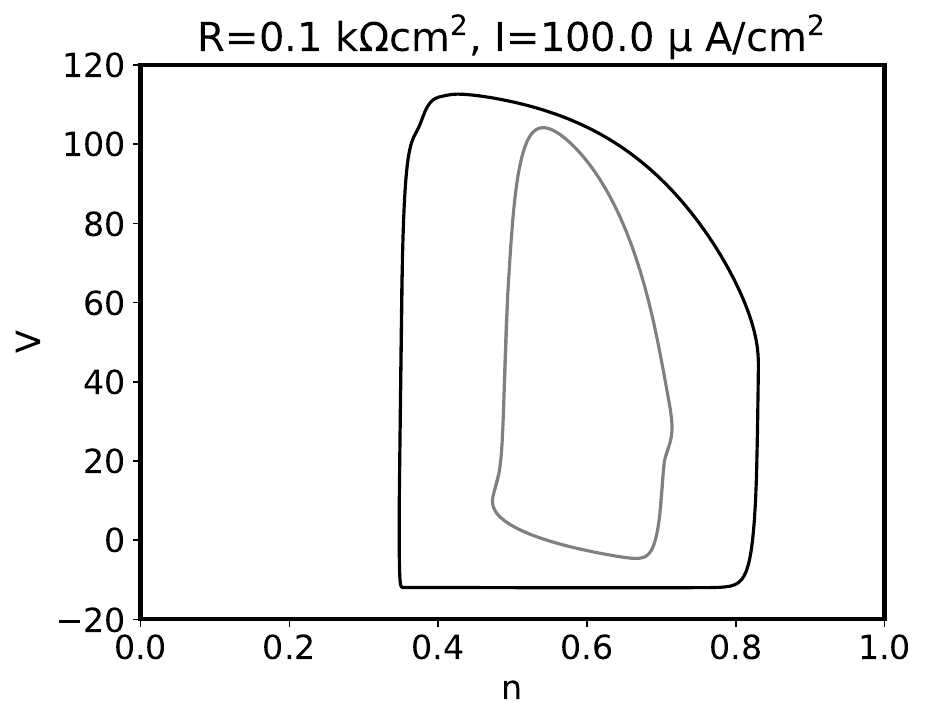}
\includegraphics[width=0.23\textwidth]{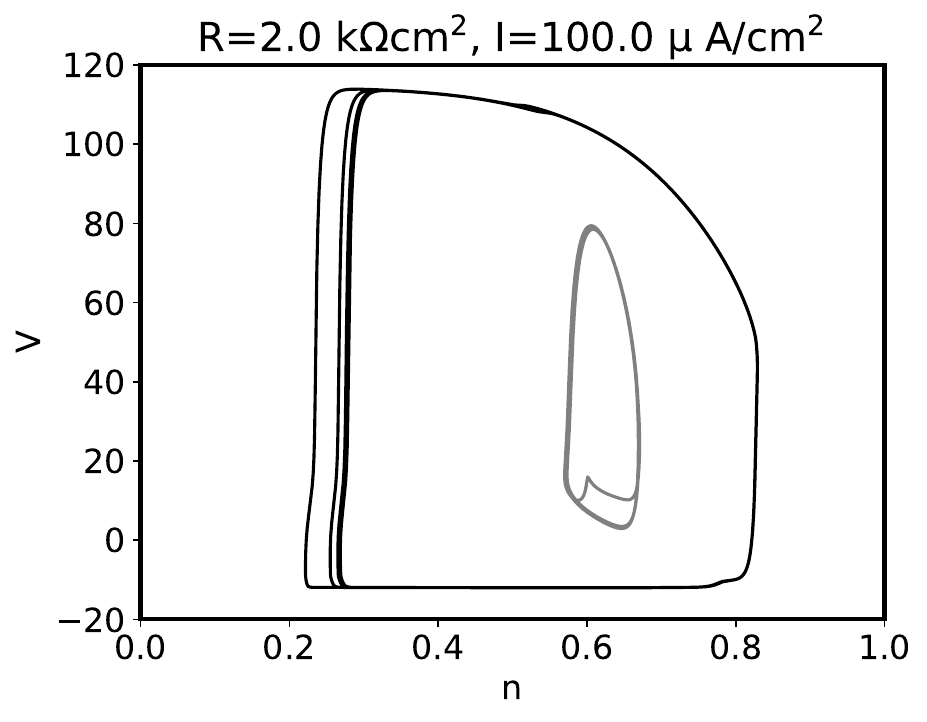}
\caption{Phase space orbits at the soma  ($N=0$, in grey) and at the middle of the axon ($x=100$, in black), obtained during the time interval $[500,2000]$~ms, with the 2D reduced HH model \eqref{propagation}.}
 \label{fig:fig10}
\end{figure}

In Figure~\ref{fig:fig10}, we depict the phase space orbits at the soma and the axon's middle for two resistivity values $R$, showing the difference between the dynamics in these axon regions.

The action potential spikes propagate at approximately constant speeds in the periodic regime. As in the 4D model, the speed of first action potential spikes are well approximated by the function $v_{\hbox{\tiny 2D}}(R,C_m=1)=4.43 \frac{1}{R^{0.66}}$, for $I=100$ and $R\in[0.01,4.6]$. 

The width of the spiky action potential signals depends on the resistivity of the axon (Figure~\ref{fig:fig9}). To quantify this effect, we calculate the width at half height, $w$, of the first action potential spike. For $I=100$ and $R\in[0.01,4.6]$, we obtain $w_{\hbox{\tiny 2D}}(R,C_m=1)=9.0 /{R^{0.76}}$~mm.

\begin{figure}[!htb]
  \centering
  \includegraphics[width=0.25\textwidth]{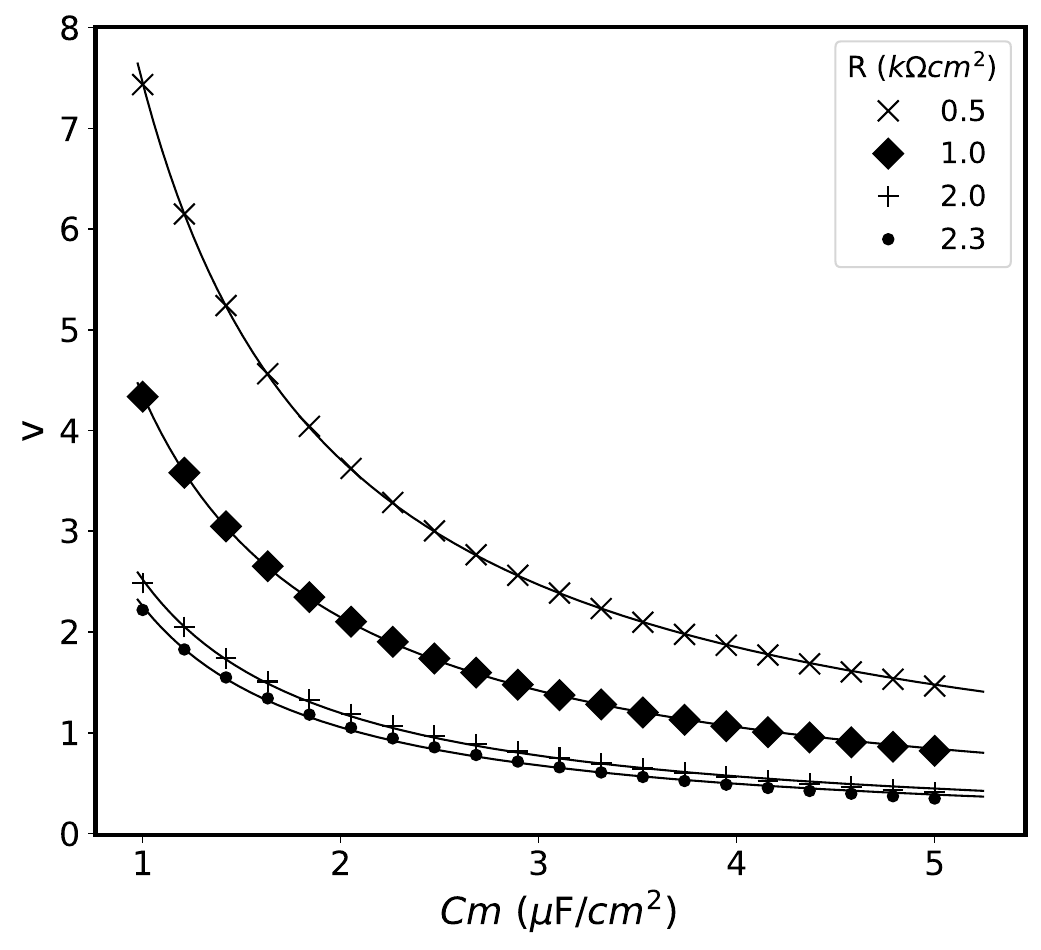}
  \caption{Relation between the propagation velocity $v$ and the membrane capacitance $C_m$ for several resistance values, calculated with the 2D reduced model.  The black lines show the fit functions  $v = \alpha_1(R)/C_m^{\beta_1(R)}$, where the parameter values for $\alpha_1(R)$ and $\beta_1(R)$ are shown in the Table \ref{tab:v(cm)}.}
  \label{fig:fig11}
\end{figure}

To examine the influence of cell capacitance on the propagation velocity of action potentials, we conducted additional simulations with various values of $C_m$ ranging from $1$ to $5$~$\mu$F/cm$^2$, while keeping some of the previously used resistance values.
We employed a fit function of $v = \alpha_1/C_m^{\beta_1}$  with an expected value of $\beta_1 = 1$, as suggested by \cite{KS}. In fact, this is justified by the change to the new time scale $\tau=t/C_m$ in equation \eqref{eq-HH}. The specific values of $ \alpha_1(R)$ and $\beta_1(R)$ obtained for different intracellular resistances are listed in Table~\ref{tab:v(cm)}, and the resulting fit functions, as well as the fitted points, are depicted in Figure \ref{fig:fig11}.

\begin{table} 
\caption{Results of fitting the data points $v(C_m)$ to the function  $v = \alpha_1(R)/C_m^{\beta_1(R)}$, for various intracellular resistances.}\label{tab:v(cm)}
  \begin{tabular}{lcccccc}
    \hline
    R $(k\Omega cm^2$)  & 0.5 & 1.0 & 2.0 & 2.3   \\
  \hline
  $ \alpha_1(R)$  & 7.45 & 4.36 & 2.52 & 2.26   \\
    $\beta_1(R)$  & 1.01 & 1.02 & 1.08  & 1.10  \\
  \hline
  \end{tabular}
\end{table}

From Table \ref{tab:v(cm)}, we conclude that $\beta_1$ is independent of the resistance $R$. The parameter $\alpha_1$ fits well with the resistivity, and we obtained $\alpha_1=4.34/R^{0.78}$. 
Merging  all these fits, it follows that the action potential velocities are well approximated by
$$
v_{\hbox{\tiny 2D}}(R, C_m)=\alpha \frac{1}{C_m R^{\beta}}
$$
where $\alpha$ and $\beta$ are constants and $\beta <1$.

\subsection{Direction of propagation of action potential signals}

We now examine how the axon responds to a current signal, either from a patch clamp experiment injected at a specific position $0<x_e<L$ along the axon's length or from a current arriving at a branching point of the axon.

\begin{figure}
\centering
\includegraphics[width=0.49\textwidth]{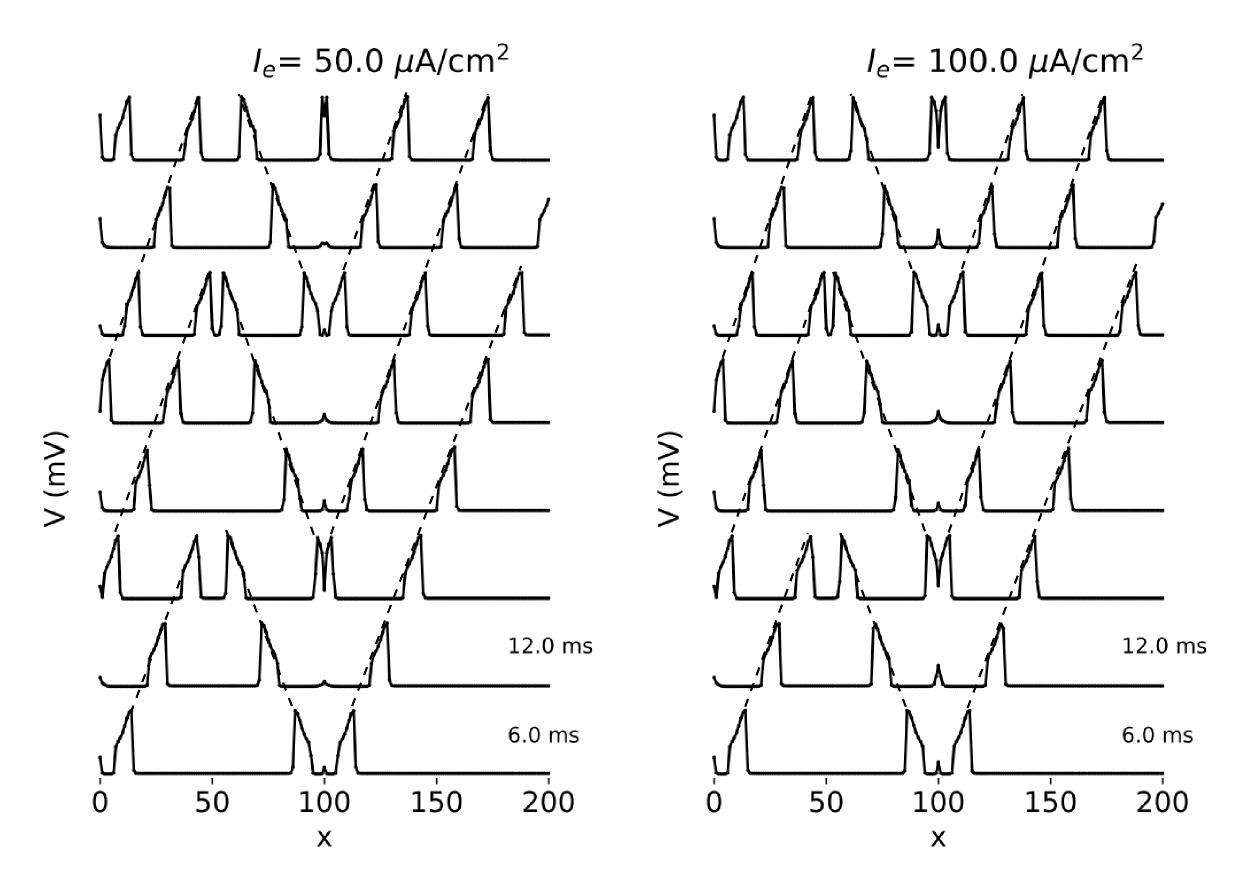}
 \caption{
 Propagation of the transmembrane potential along the axon with a constant stimulus at the soma with $I=100$~$\mu$A/cm$^2$,  and second constant stimuli at the axon position $x_e=100$, with $I_e=50$~$\mu$A/cm$^2$ and $I_e=100$~$\mu$A/cm$^2$. The axon has $N=200$ segments with lengths $\ell=1$~mm, and the vertical axis range is $[-20,125]$~mV. The other parameters of the simulation are  $C_m = 1$~$\mu F/$cm$^2$ and $R=2$~k$\Omega$cm$^2$. 
The dashed lines are the characteristic curves of the action potential spikes with a slope proportional to the speed of the spikes. The signal injected at $x_e=100$ annihilates the signal arriving from the soma. The signal that arrives at the synaptic region of the axon is generated in the middle of the axon.}
 \label{fig:fig12}
\end{figure}

In Figure~\ref{fig:fig12}, we depict the temporal evolution of the axon's membrane potential in response to a constant stimulus $I=100$~$\mu$A/cm$^2$ originating from the soma, along with localised stimuli at $x_e=100$ with $I_e = 50$ (left) and $I_e = 100$ (right). We considered that $R=2$~k$\Omega$cm$^2$ and $C_m = 1$~$\mu F/$cm$^2$.

In these simulations, two action potential train spikes are produced at $x_e=100$, propagating in opposite directions of the axon. One of the spikes annihilates the membrane potential response generated at the soma, while the other action potential spike produced at the branching point propagates towards the presynaptic region of the axon. These effects were also found in the 4D HH model in extended domains \cite{CD2}.

\section{Conclusions}\label{sec4}
\label{sec:concl}

We successfully reduced the 4D HH model to novel 3D and 2D models while preserving its dynamical and electrophysiological properties and maintaining the same Bautin bifurcation scenario \cite{CD1}. This approach differs from other heuristic methods outlined by \cite{FH} and \cite{Abb}. A crucial step in this reduction involves substituting the sodium gating variable $h$ in the 4D HH model with $c(I)-n$. This relationship between the gating variables $h$ and $n$ has been documented in the literature under the hypothesis that $c$ is a constant independent of the stimulus at the soma, \cite{FH}, \cite{Rin2},  \cite{KS} and \cite{Wan}. This underscores the interconnectedness between sodium and potassium voltage-gated channels, ultimately described by Na$^+$/K$^+$-ATPase pumps distributed along axons \cite{Bea}. This method successfully derived a 2D Hodgkin-Huxley-type model, which simplified the analysis of action potential dynamics.


We found that the propagation of action potentials occurs within a specific range of axon resistivity, $R\in [R_{m0},R_{M0}]$. Beyond these limits, the axon remains in a steady state regardless of the soma's (current) signal intensity. Within the resistivity interval $[R_{m0},R_{M0}]$, there is a narrower interval $[R_{m1},R_{M1}]\subset [R_{m0},R_{M0}]$ where periodic or nearly periodic action potentials are generated at each spatial position along the axon. For $R\in [R_{m0},R_{M0}]-[R_{m1},R_{M1}]$, the sustained current originating from the soma generates a solitary action potential spike or soliton that propagates along the axon. Upon collision at the presynaptic region of the axon, the axon returns to a steady state.

For the $4D$, $3D$ and $2D$ HH type models, the propagation speed of the action response is well-described by the function
\[
v(R, C_m)\simeq \alpha  \frac{1}{C_m R^{\beta}}:=\gamma D^{\beta}
\]
where $\alpha $ and $\beta <1$ are positive constants independent of the soma stimulus intensity. $D$ is the diffusion coefficient of the axon, while $\gamma$ is a constant determined by the axon's membrane electric properties. For the three models, we have obtained $\beta=0.55$ (4D), $\beta=0.52$ (3D) and $\beta=0.66$ (2D), and  the width of the action potential spikes also depends on 
the resistivity of the axon with $w(R, C_m=1)=\alpha_2 /R^{\beta_2}$, where $\alpha_2$ and $\beta_2$ are positive constants.

The simplicity of using the 2D simplified model to represent real neurons accurately may lead to more precise simulations of neural circuits and networks. 

\section*{Data availability statement}
No data is associated with the manuscript.

\section*{Competing interests}
  The authors declare that they have no competing interests.

\end{document}